\documentclass[a4paper]{jpconf}
\usepackage{color,amsmath,amssymb}
\usepackage[]{graphicx}
\usepackage{float}
\usepackage{placeins}
\usepackage[font=scriptsize,labelfont=bf]{caption}

\begin{document}
\title{$T$-Matrix Approach to Strongly Coupled QGP}

\author{Shuai Y.F. Liu, Ralf Rapp}
\address{Cyclotron Institute and Department of	Physics\&Astronomy, Texas A\&M University, 
	College Station, TX 77843-3366, USA}
\ead{lshphy@gmail.com, rapp@comp.tamu.edu}

\begin{abstract}
Based on a thermodynamic $T$-matrix approach we extract the potential $V$ between
two static charges in the quark-gluon plasma (QGP) from fits to the pertinent lattice-QCD 
free energy. With suitable relativistic corrections we utilize this new potential to
compute heavy-quark transport coefficients and compare the results to previous calculations
using either $F$ or $U$ as potential. We then discuss a generalization of the $T$-matrix 
re-summation to a ``matrix $\log$" re-summation of $t$-channel diagrams for the grand partition
function of the QGP in the Luttinger-Ward skeleton diagram formalism. 
With $V$ as a non-perturbative driving kernel in the light-parton sector, we obtain the QGP 
equation of state from fits to lattice-QCD data. The resulting light-parton spectral 
functions are characterized by large thermal widths at small momenta, indicating the
dissolution of quasi-particles in a strongly coupled QGP.

\end{abstract}

\section{Introduction}
The large suppression and elliptic flow of heavy-flavor (HF) spectra in heavy-ion collisions (HICs) 
at RHIC and LHC indicate that heavy quarks couple strongly to the quark-gluon plasma (QGP), 
see, {\it e.g.},~Ref.~\cite{Prino:2016cni} for a recent review. 
To address this problem non-perturbatively, we have been developing a many-body $T$-matrix 
approach that includes fundamental features of the QCD force constrained by lattice QCD 
(lQCD)~\cite{vanHees:2007me,Riek:2010fk}. One of the debated issues in such approaches is 
precisely which quantity is the appropriate heavy-quark (HQ) potential, $V$. This may depend 
on the definition of the potential in a given approach. In practical applications, one often 
employs the free ($F$) and internal ($U$) energies, which are directly available from lQCD, as 
limiting cases. The former gives rise to a rather weak in-medium force, while for the latter 
it is substantially stronger, which is, in fact, preferred by phenomenological applications, 
{\it e.g.}, to HF diffusion in HICs~\cite{He:2014cla}. However, a 
more rigorous and quantitative determination of the potential is desirable. Toward this end, 
in recent work~\cite{Liu:2015ypa}, we have defined the potential as a driving kernel in a 
thermodynamic $T$-matrix equation, derived the pertinent free energy and, through fits to lQCD 
data for $F$, extracted the quantitative form of the underlying $V$. 

In the present paper, we will briefly review this work and then discuss first applications
to the HQ transport coefficients and, by extending it to the light sector, to
the equation of state (EoS) of the QGP. 

\section{Heavy-Quark Potential and Transport Coefficients}
\begin{figure}[!t]
\begin{center}
	\begin{minipage}{28pc}
		\includegraphics[width=14pc]{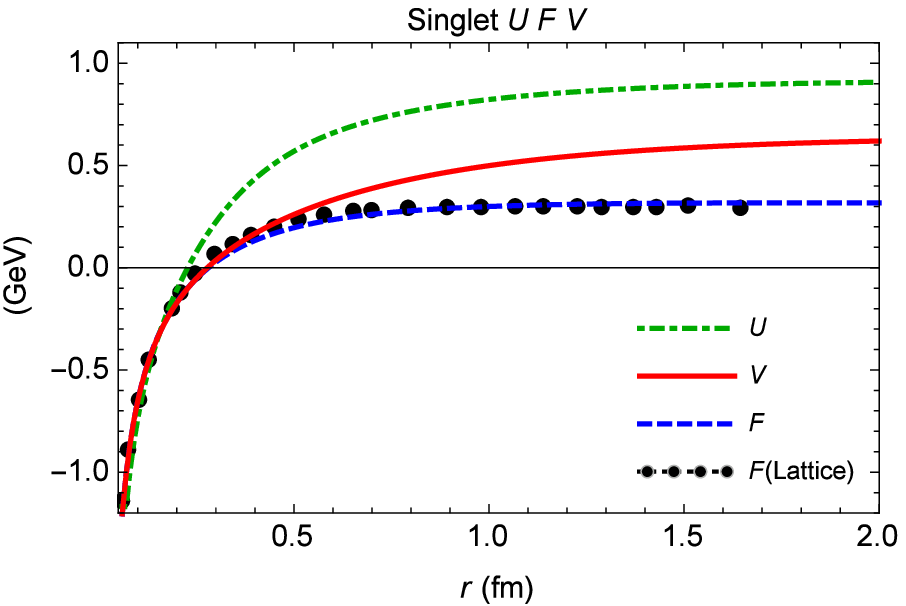}\hspace{0.5cm}
		\includegraphics[width=14pc]{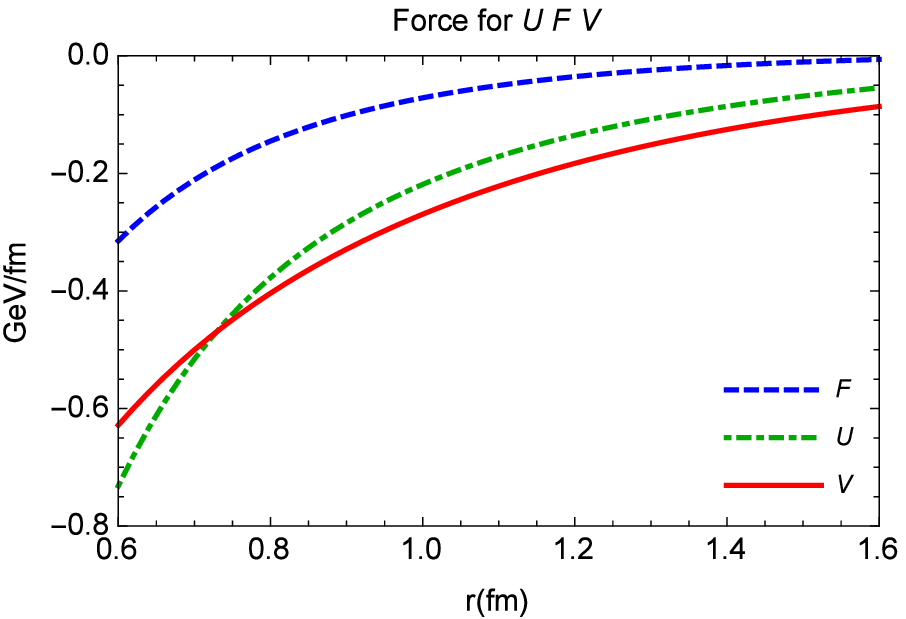}
		\caption{Color-singlet $U$, $F$ and $V$ (left) and corresponding forces (right) at $T$=0.24\,GeV.}
		\label{fig_pot}
	\end{minipage} 
\end{center}
\end{figure}

In Ref.~\cite{Liu:2015ypa}, the thermodynamic $T$-matrix approach has been used as a starting
point for defining the HQ potential, $V$. The resulting expression for the static HQ free energy, $F$, was obtained as
\begin{align*}
F_{}(r)=-\frac{1}{\beta} \ln\left[\int^{\infty}_{-\infty}d\omega \frac{-1}{\pi}\text{Im}\left[\frac{1}{[\hat{G} (E+i\epsilon)]^{-1}-V(E+i\epsilon,r) }\right] e^{-\beta\omega}\right]
\end{align*}
where $\hat{G}$ is the non-interacting 2-body Green's function including in-medium HQ self-energies.
A 4-parameter ansatz for $V$ was made consisting of screened Cornell and string terms. 
The resulting fit to lQCD data for $ F $~\cite{kaczmarek2005static,kaczmarek2007screening,Petreczky:2004pz} 
is shown in Fig.~\ref{fig_pot} left, along with the underlying potential and the ensuing internal 
energy, $U$. While the magnitude of $V$ lies between $F$ and $U$, it turns out that $V$ 
actually has the largest force at large distance, for $r\ge 0.7$\,fm (Fig.~\ref{fig_pot} right). 
This will allow heavy quarks to interact with more medium partons at a time, with important 
consequences for the transport coefficients.

\begin{figure}[!h]
\begin{center}
	\begin{minipage}{30pc}
		\includegraphics[width=15pc]{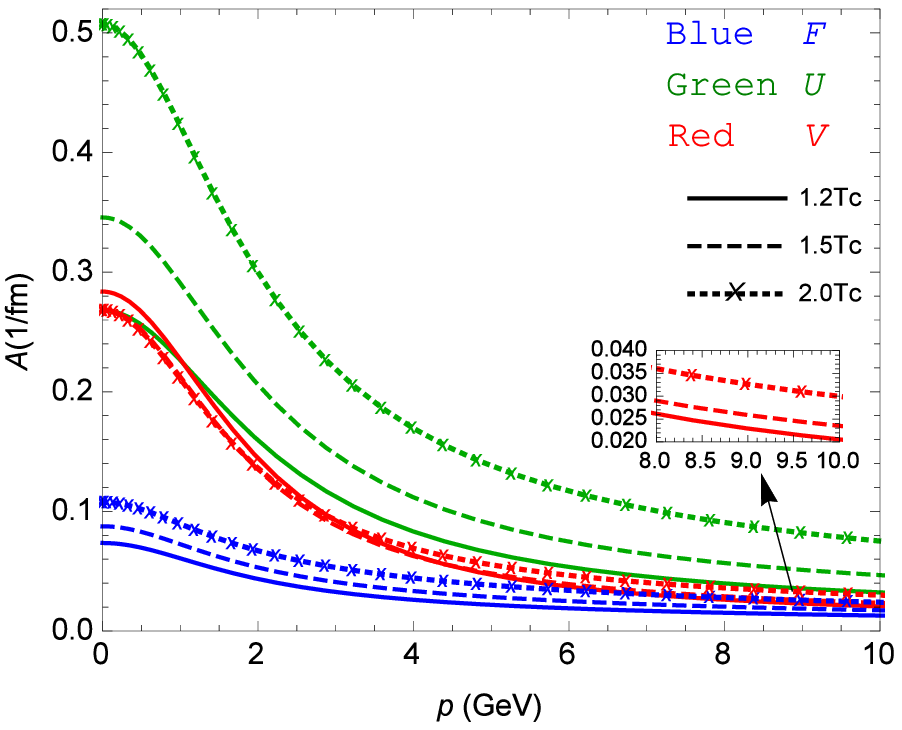}\hspace{0.5cm}
		\includegraphics[width=15pc]{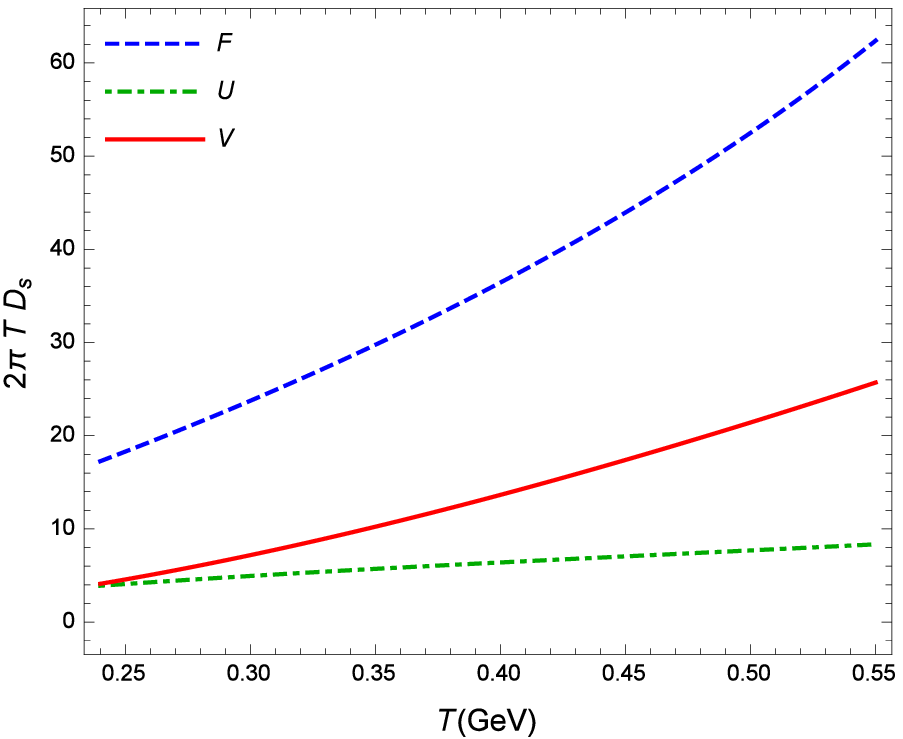}
	\caption{Drag (left) and spatial diffusion coefficients (right) calculated for $F$, $U$ and $V$.}
	\label{fig_coeff}
	\end{minipage}
\end{center} 
\end{figure}

Next, we use $V$ in the $T$-matrix formalism~\cite{vanHees:2007me,Riek:2010fk,Huggins:2012dj} 
to calculate the thermal HQ relaxation rate within a Fokker-Planck framework~\cite{Svetitsky:1987gq}
assuming either $F$, $U$ or $V$ as potential, see Fig.~\ref{fig_coeff} left. At low momenta, the rate 
from $V$ is the largest, due to the long-range force in the potential as mentioned above. Quite remarkably,
the temperature ordering is reversed, due to large contributions from the string term. 
A schematic dimensional analysis suggests the cross section for the string term to be proportional to 
$\sigma/T^4$, where $\sigma$ is the string tension. This low-$T$ enhancement overcomes the decrease 
in density ($\sim T^3$) when $T$ decreases, thus leading to an increased thermalization rate.
With increasing momentum, the strong coupling transitions into a weakly coupled regime driven
by short-range color-Coulomb interactions at high momentum, recovering a normal temperature hierarchy.  
The resulting spatial diffusion coefficients, scaled by the thermal wavelength, are displayed in
Fig.~\ref{fig_coeff} right.

\section{Equation of State}
\begin{figure}[!t]
\begin{center}
		\begin{minipage}{14pc}
			\includegraphics[width=15pc]{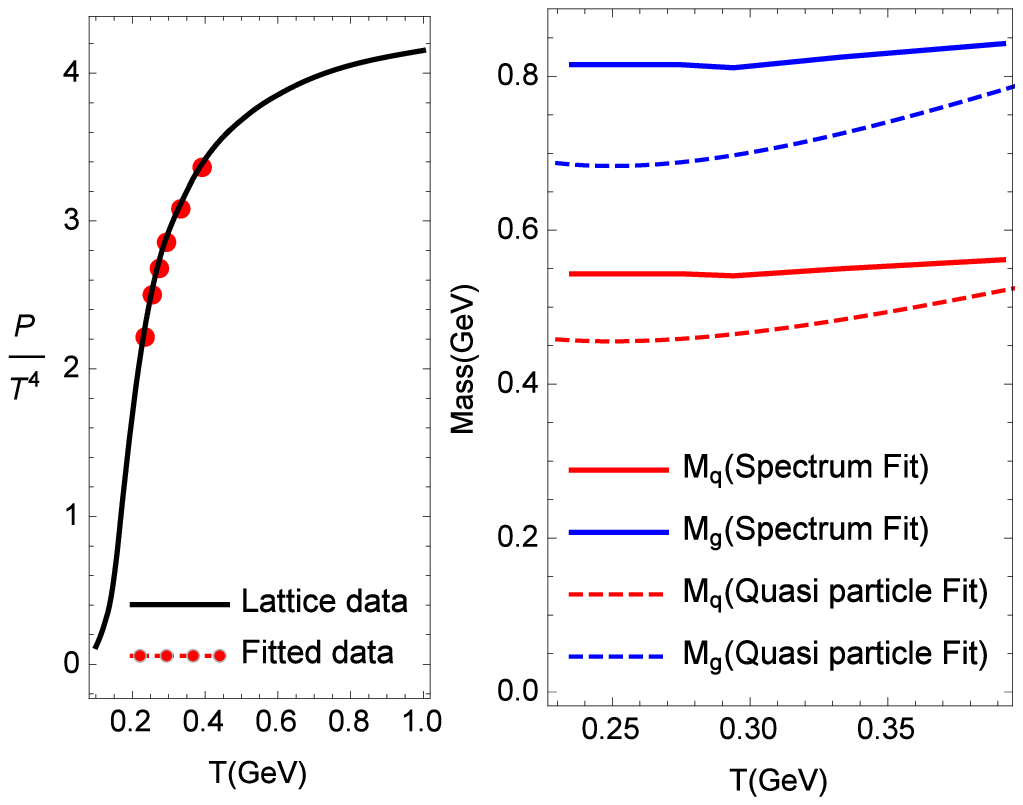}
			\caption{Fit to the lattice-QCD EoS (left) and the fitted quark and gluon masses (right).}
\label{fig_eos}
		\end{minipage}\hspace{2pc}%
		\begin{minipage}{19pc}
		\includegraphics[width=18pc]{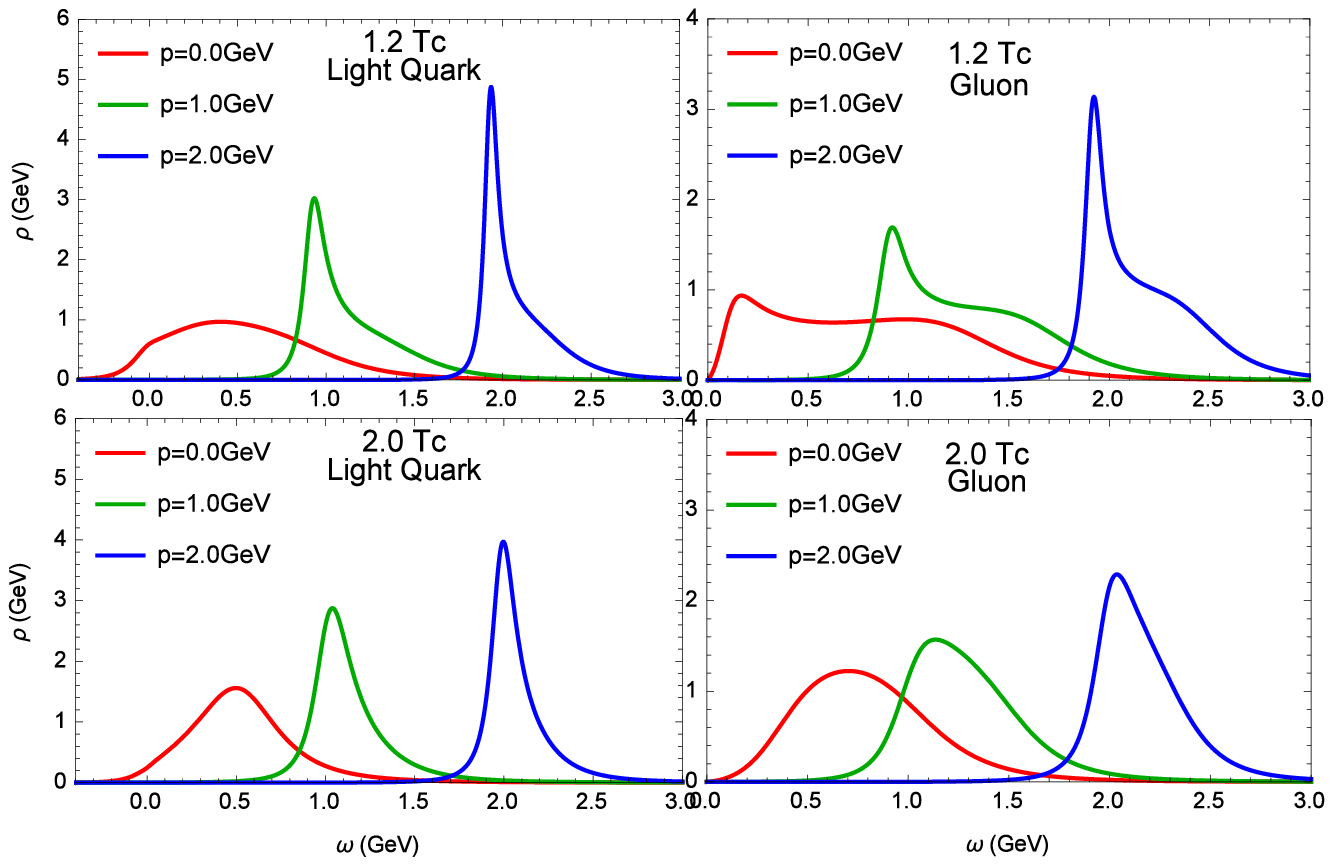}
		\caption{Light-parton spectral functions at different temperatures and momenta ($T_c$=196\,MeV).}
\label{fig_spec}
		\end{minipage}
\end{center}
\end{figure}

The large drag coefficient obtained from $V$ translates into large HQ self-energies, with pertinent
widths of order 0.5-1\,GeV. To investigate the impact of the strong coupling on the ``light"-parton 
sector, we have developed a generalized $T$-matrix formalism to compute the EoS of QGP. Specifically, 
we re-sum all $t$-channel diagrams in the Luttinger-Ward skeleton expansion~\cite{PhysRev.118.1417,Baym:1962sx} through a novel
matrix ``$ \log $" technique which allows us to include both single-parton and resonance (bound-state) 
contributions on the same footing, with self-consistent thermal widths.
More explicitly, the Luttinger-Ward functional can be schematically written as
\begin{align}
\Omega (T)=\underset{s,c,f}{\sum }\pm\int \tilde{d^4p} \left[\left[\ln \left(-G^{-1}\right)+\Sigma G\right]
-\underset{\nu}{\sum }\frac{1}{2\nu}\Sigma _\nu G\right] \ .
\end{align}
The single-particle self-energy in $T$-matrix approximation can be written as
\begin{align}
\Sigma &=\underset{s,c,f}{\sum }\int \tilde{d^4p}TG=\underset{s,c,f}{\sum }\int \tilde{d^4p}\left\{V+V\hat{G}V+\ldots
	+V\hat{G}V\hat{G}\ldots .V\right\}G\\\nonumber
&=\underset{s,c,f}{\sum }\int \tilde{d^4p}V\left(1-\hat{G}V\right)^{-1}G \ ,  
\end{align}
where the re-summation of the ladder series is carried out through matrix inversion. 
In analogy, we identify the series for the skeleton expansion via a natural-$\log $ function ($\ln$) 
which can be resummed into a ``matrix-$\log$",
\begin{align}
\underset{\nu}{\sum }\frac{1}{2\nu}\Sigma _\nu&=\frac{1}{2}\underset{s,c,f}{\sum }\int \tilde{d^4p}\left\{V+\frac{1}{2}V\hat{G}V+\ldots
	+\frac{1}{v}V\hat{G}V\hat{G}\ldots .V\right\}G
\nonumber\\
&=\frac{1}{2}\underset{s,c,f}{\sum }\int \tilde{d^4p}\left\{-\hat{G}^{-1}\ln \left(1-\hat{G}V\right)\right\}G \ .
\end{align}
Applying this expression with the in-medium $V$ as the potential, we calculate 
the EoS of the QGP {\em self-consistently}. Similar to quasi-particle approaches~\cite{Plumari:2011mk}, 
we employ the quark and gluon masses as two fit parameters to the EoS of lQCD~\cite{Borsanyi:2010cj}, see
Fig.~\ref{fig_eos}. The self-consistent formalism generates self-energies with non-trivial energy-momentum 
dependencies, in particular thermal widths. The latter turn out to be very large at low parton momenta, 
of order 1(0.5)\,GeV at $T$=1.2(2.0)$T_c$, see Fig.~\ref{fig_spec}. This implies the loss of well-defined 
quasi-particle excitations as $T_c$ is approached from above, while at higher $T$ and/or 3-momenta ($p$), 
quasi-particle peaks are resurrected in the spectral functions. Our calculations thus suggest a rich 
structure of the QGP, with strong coupling for soft modes and a transition to a quasi-particle like, more 
weakly coupled QGP for harder modes.  Again, we believe that these structures are critically driven 
by  ``cross sections" (although we do not use this notion) 
for the string term ($\sim$$\sigma/T^4$) which dominate over the reduction in density ($\sim$$T^3$) as $T$ is lowered, 
rendering a more strongly coupled medium. This feature is intimately related to the $T$ and $p$ dependencies 
of the HQ transport coefficients discussed in the previous section.
We note that we currently encounter instabilities of the self-consistent iteration procedure 
when computing gluon self-energies from the $T$-matrix close to $T_c$. These instabilities require further studies in the future, 
together with calculations of susceptibilities at finite chemical potential. This may help us to understand 
the phase structure generated by the $T$-matrix approach and its relation to QCD phase structure. 

\section{Conclusion and Perspective}
We have employed a thermodynamic $T$-matrix approach with non-perturbative interactions
to study HQ transport and the EoS of the QGP. We have extracted a HQ potential $V$ by
fitting the HQ free energy to lQCD data, and compared HQ transport coefficients with different
driving kernels ($F$, $U$ and $V$) for heavy-light interactions. We find that the low-momentum drag 
coefficient is largest for $V$, due to a long-range remnant of the confining force, with a nontrivial
temperature dependence. We have extended the $T$-matrix to the light sector to study the EoS 
of the QGP. Within a self-consistent skeleton formalism, we find light-parton spectral 
functions with large widths at low momenta and temperatures, decreasing with temperature. 
This behavior directly reflects the properties of the HQ drag coefficients. Large values for parton
widths and HQ drag coefficients suggest the QGP to be strongly coupled, where soft modes are no
longer quasi-particles while heavy quarks may still serve as a suitable probe.
In the future, we will work toward a unified non-perturbative $T$-matrix framework for 
heavy and light sectors that can simultaneously describe various lQCD data such as the EoS, 
susceptibilities, correlator ratios, and static free energies~\cite{Liu:2015ypa}.
We believe that the variety of lQCD data can help to constrain the 
uncertainties of the framework so that we can perform controlled calculations of transport 
coefficients that directly link to experimental observables in HICs.

%
%

\section*{Acknowledgments}
This work is supported by the U.S.~NSF through grant no. PHY-1614484.
\section*{References}
\bibliographystyle{iopart-num}

\bibliography{refc}

\end{document}